# HEAT LOAD AND STRESS STUDIES FOR AN DESIGN OF THE PHOTON COLLIMATOR FOR THE ILC POSITRON SOURCE

F. STAUFENBIEL[*,1], O.S. ADEYEMI[2], V. KOVALENKO[2], G. MOORTGAT-PICK[2,3], L. MALYSHEVA[2], S. RIEMANN[1], A. USHAKOV[2]

[1] *DESY, Platanenallee 6, Zeuthen, D-15738, Germany*

[2] *II.Insitut für Theor. Physik, University of Hamburg, Luruper Chaussee 149, Hamburg, D-22761, Germany*

[3] *DESY, Notkestr.85, Hamburg, D-22761, Germany*

\* E-mail: friedrich.staufenbiel@desy.de

The ILC baseline design for the positron source is based on radiation from a helical undulator to produce positrons in a thin target. Since the photon beam created in the helical undulator is circularly polarized, the generated positron beam is longitudinally polarized. Using a photon collimator upstream the positron target the positron polarization can be enhanced. However, the photon beam intensity yields a huge thermal load in the collimator material. In this paper the thermal load and heat dissipation in the photon collimator is discussed and design solutions are suggested.

## 1 Introduction

The ILC project proposes a high-luminosity electron-positron collider with cms-energies up to 1TeV [1]. The electron beam will have a polarization larger than 80%. The positron source design is based on a helical undulator [2] passed by the high-energy electron beam to radiate circularly polarized photons. The photon beam hits a thin Ti-alloy target and produces pairs of longitudinally polarized electrons and positrons. A photon collimator upstream the target cuts that part of the photon beam with the lower average polarization. With increasing photon beam collimation the energy deposition in the collimator material increases. Due to the time structure of the intense photon beam the peak energy deposition density (PEDD) could exceed the limits accepted by the collimator material. The material limits are given by the elastic yield strength (corresponding to the thermal activation energy of diffusion) and by the fatique stress (corresponding to the thermal recrystallization energy) over a long-term running. The limits can be estimated by temperature loads with 70% and 40% of the material melting points, respectively (e.g. see [3]). Therefore, it is necessary to find a proper design and material choice to prevent a collimator system breakdown. This paper presents the results of a design study for a photon collimator system for the ILC positron source. This study is based on an undulator design suggested for cms-energies up to 500GeV. In section 2 the parameters of the positron production system are given. In section 3 the heat load, the choice of the collimator material are discussed and in section 4 its dissipation; the cooling requirements are included. The last section summarizes and presents an outlook.

## 2   Production of polarized positrons by helical undulator radiation

The parameters of the electron drive beam are given in reference [1]. Here we consider an electron beam energy of about 250GeV, $2 \times 10^{10}$ electrons per bunch and 2625 bunches per train with a train length of about 969µs and 5Hz repetition rate as suggested in the SB2009 proposal [4]. Our results correspond to the parameters presented in the Technical Progress Report 2011 [5] by scaling.

In order to produce circularly polarized photons the electrons pass a helical undulator with K=0.92, $\lambda_0$=11.5mm, located at a distance of about 500m to the positron target and photon collimator respectively. The effective undulator length is $L_{und}$ =45m to generate the required number of positrons with a 250GeV electron drive beam. The resulting average photon beam power is $P_\gamma$=164kW.

The opening angle of the radiated photon beam is determined by the energy of the electron beam; it is proportional to $1/\gamma$. The opening angle of the higher harmonics cone is $K/\gamma$.

The degree of photon polarization depends on the emission angle and on the fractional energy of the photons (see [6]). By cutting the outer part of the radial symmetric beam, the polarization increases by contemporaneous decreasing positron yield. For the above mentioned parameter set, Table 1 shows the degree of positron polarization and the production yield for three selected collimator apertures (see also [7]).

The intensity of the undulator radiation has the maximum around the beam axis. For the given parameters, a positron polarization of 60% can be achieved for photon beam radii collimated to r=1.0mm. With larger collimation radii the positron polarization approaches about 27% for r≥3.0mm (see [8]). However, to provide the required number of positrons also for high values of polarization, a more intense photon beam is required resulting in a substantially increased heat load in the collimator material due to the increased photon beam power. For example, for 60% positron polarization the photon beam power is twice as much as for 27%.

Tab.1. Polarization degree and positron yield for different collimator apertures.

| collimator aperture | $P_{e+}$ | $e^+$ Yield |
|---|---|---|
| no collimator | 27% | 100% |
| 2.0 mm | 35% | 91% |
| 1.4 mm | 47% | 73% |
| 1.0 mm | 60% | 50% |

## 3  Heat load studies for photon collimators

The aperture of the collimator determines the average polarization of the photon beam and hence, of the positrons produced in the titanium target and captured [8]. The heat load in the collimator and target is calculated using the FLUKA Monte Carlo code for particle tracking and particle interactions with matter [9]. By means of this simulation tool the optimization of the photon collimator design is done by quantifying the energy deposition in the collimator materials and selecting them corresponding to the tolerable increase of temperature and material stress, respectively [10]. The temperature rise is calculated by

$$\Delta T = \frac{\Delta Q}{m \cdot c_v} \qquad (1)$$

where $\Delta Q$ is the energy deposition in the material (given in [J]), m the mass and $c_v$ is the heat capacity (unit [J/kg/K]). The maximal temperature rise corresponds to the maximal PEDD.

### 3.1  *Previous photon collimator design*

So far, the design for the high power photon collimator was based on electron beam energies of 150GeV which leads to photon energies up to 10MeV for the first harmonic radiation. The collimator had a cylindrical shape and was segmented in a first part made of graphite and a second part made of tungsten [11], each with a length of z=9cm [12]. Providing the required numbers of positrons for the parameter set given above, an instantaneous temperature rise up to $\Delta T=1800K$ is expected in the tungsten part for an aperture of r=2.0mm. For an aperture of r=1.0mm the temperature rises up to $\Delta T=14000K$. Therefore, the designs presented in [11,12] are not appropriate for the considered ILC beam parameter set.

### 3.2 *Moveable multistage collimators*

To achieve more flexibility in the polarization and yield manipulation for the positron beam, a moveable multistage collimator system with attenuating apertures is proposed as shown in Figure 1. The designs of the second and the third collimator take into account the collimation from the previous collimator in order to keep the lengths of the whole device as short as possible. The following calculations and simulations correspond to such system.

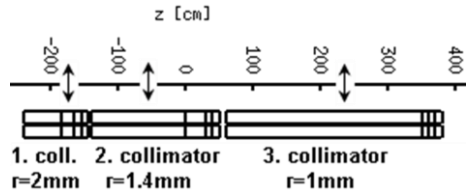

Fig.1. Moveable multistage photon collimators enable a flexible choice of the positron polarization (e.g., 27%, 35%, 47%, 60%). For this solution the required space in z direction is less than 7m.

So far, a design of moveable collimator segments each consisting of one material has been considered. The fabrication of these segments with small aperture including cooling channels has not been regarded; probably the longer collimator components will consist of partitioned segments. The proper alignment of a collimator with small aperture segments requires special care to obtain the desired reproducibility of polarization.

### 3.3 *Optimized collimator design*

The main fraction of energy is absorbed in the first part of the collimator. A low-Z material - pyrolytic graphite - is chosen. It's evaporation point scores up to 3650ºC without a liquid phase [13]. In addition, pyrolytic graphite is very resistant against particle evaporation by energy impact. The material is strong anisotropic in the (xy) plane (basal direction) and the (z) direction. The thermal conductivity is a factor of 200 higher in the basal direction and has a very low thermal expansion coefficient [14]. However, due to the high radiation length of $X_0 \approx 19$cm, a long collimator would be needed to absorb the unwanted part of the photon beam. In order to distribute the energy deposition over a large range in the collimator material and to keep the collimator short, a proper medium-Z (or high-Z) material with smaller radiation length has to follow the graphite segment. Figures 2 to 4 show the simulated energy deposition distributions in collimators for three different apertures; the collimators are assembled with structures of pyrolytic graphite, titanium, iron and tungsten.

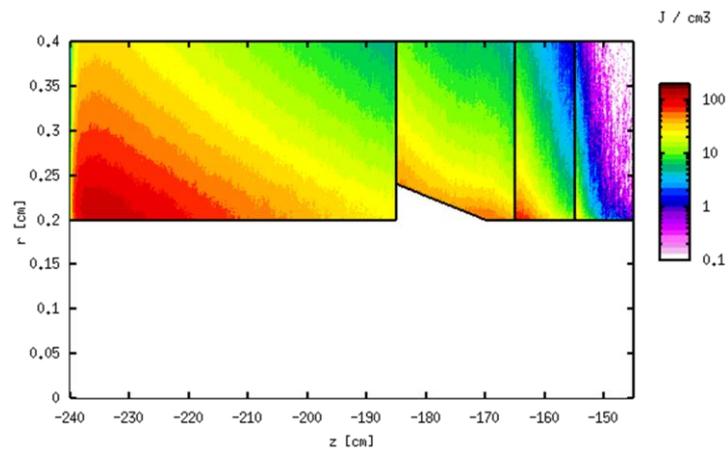

Fig.2. FLUKA simulation of the energy distribution deposited by the circularly polarized photon beam in a collimator composed of pyrolytic graphite, titanium, iron and tungsten, with an aperture 2.0mm radius and a length of z=95cm.

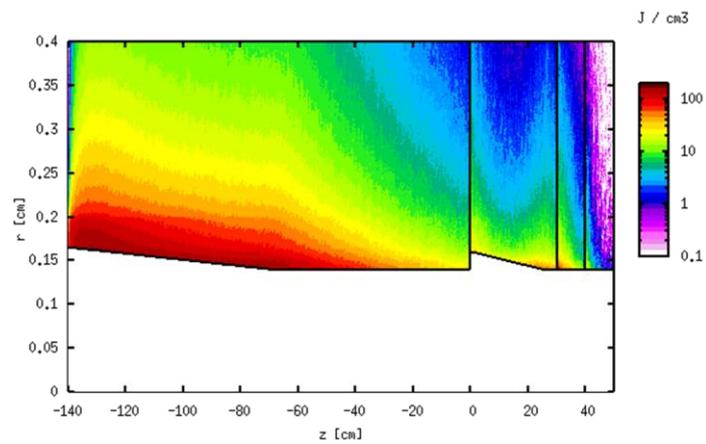

Fig.3. FLUKA simulation of the energy distribution deposited by the circularly polarized photon beam in a collimator composed of pyrolytic graphite, titanium, iron and tungsten, with an aperture 1.4mm radius and a length of z=190cm.

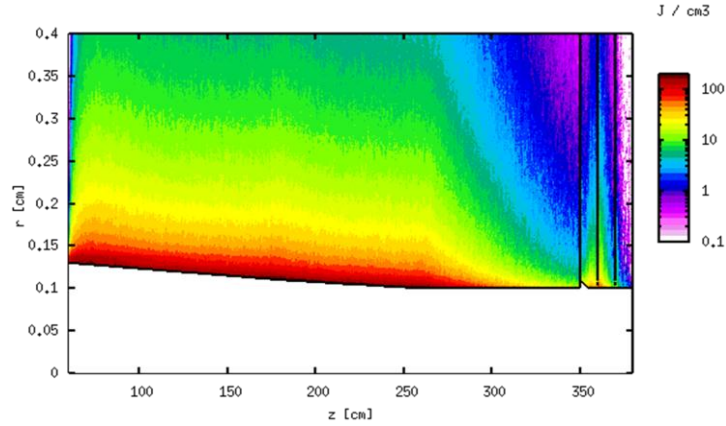

Fig.4. FLUKA simulation of the energy distribution deposited by the circularly polarized photon beam in a collimator composed of pyrolytic graphite, titanium, iron and tungsten, with an aperture 1.0mm radius and a length of z=320cm.

All unwanted photons and secondary particles are absorbed in the collimator. The lengths of the parts are optimized to lower the energy deposited in the following higher Z material to an acceptable level. The length of the last very high Z material tungsten is optimized in order to reduce the exiting shower particles below 0.1% of the absorbed photon power. For the collimator design presented in Figures 2-4 more than 99.9% of the unwanted part of the photon beam is absorbed; less than 0.1% reaches the titanium target.

Table 2 shows the PEDD and the maximum temperature rise respectively for a multistage photon collimator.

Tab.2. PEDD in the components of the multistage photon collimator. The aperture radii of 2.0mm, 1.4mm and 1.0mm correspond to 35%, 47% and 60% polarization degree for the produced positron beam. The effective undulator length is 45m, and a photon yield of 1.95ph/(e- m) yields to (4.6e15)ph/train.

| Aperture r=2.0mm | $E_{max}$ per train [J/cm$^3$] | $\Delta T_{max}$ per train [K] | $\Delta T_{max}$ for $P_{e+}$=60% $L_{und}$=90m per train [K] |
|---|---|---|---|
| graphite | 111 | 70 | 140 |
| titanium | 56 | 24 | 48 |
| iron | 78 | 22 | 44 |
| tungsten | 31 | 12 | 24 |
| aperture r=1.4mm | $E_{max}$ per train [J/cm$^3$] | $\Delta T_{max}$ per train [K] | $\Delta T_{max}$ for $P_{e+}$=60% $L_{und}$=90m per train [K] |
| graphite | 159 | 100 | 200 |
| titanium | 54 | 23 | 46 |
| iron | 78 | 22 | 44 |
| tungsten | 30 | 11 | 22 |
| aperture r=1.0mm | $E_{max}$ per train [J/cm$^3$] | $\Delta T_{max}$ per train [K] | $\Delta T_{max}$ for $P_{e+}$=60% $L_{und}$=90m per train [K] |
| graphite | 176 | 110 | 220 |
| titanium | 54 | 23 | 46 |
| iron | 78 | 22 | 44 |
| tungsten | 31 | 12 | 24 |

The photon collimator decreases the positron yield as shown in Table 1. This has to be compensated by increasing the undulator length to $L_{und}=90m$ according to $N_\gamma \sim L_{und}$ in order to maintain the demanded luminosity at the IP. This results in a higher heat load which is taken into account in the last column of Table 2.

Due to the energy distribution in the photon beam the heat load on the material increases substantially with decreasing collimator radii. In order to distribute the load over a wide range of the collimator material, a conical shape at critical areas of high energy depositions is used. Therefore, the first sections of the graphite part from the second and third collimator and the first titanium part have a conical shape. The length of the graphite part increases strongly for smaller apertures.

The collimator design presented in Figures 2-4 implies a safety-factor to allow moderate average temperatures in the bulk materials and an efficient cooling. The dimensions of the collimators are summarized in the Tables 3-5.

### 3.3.1 *Dimensions of the first collimator with 2.0mm aperture*

Tab.3. Dimensions of the parts for the first collimator in the whole arrangement. The total length is about 95cm.

| collimator part | aperture $r_1,r_2$[mm] | length z [cm] |
|---|---|---|
| graphite / cyl. | 2.0 | 55 |
| titanium / cone | 2.4 / 2.0 | 15 |
| titanium / cyl. | 2.0 | 5 |
| iron / cyl. | 2.0 | 10 |
| tungsten / cyl. | 2.0 | 10 |

### 3.3.2 *Dimensions of the second collimator with 1.4mm aperture*

Tab.4. Dimensions of the parts for the second collimator in the whole arrangement. The total length is about 190cm.

| collimator part | aperture $r_1,r_2$[mm] | length z [cm] |
|---|---|---|
| graphite / cone | 1.65 / 1.4 | 70 |
| graphite / cyl. | 1.4 | 70 |
| titanium / cone | 1.6 / 1.4 | 25 |
| titanium / cyl. | 1.4 | 5 |
| iron / cyl. | 1.4 | 10 |
| tungsten / cyl. | 1.4 | 10 |

### 3.3.3 *Dimensions of the third collimator with 1.0mm aperture*

Tab.5. Dimensions of the parts for the third collimator in the whole arrangement. The total length is about 320cm.

| collimator part | aperture $r_1, r_2$ [mm] | length z [cm] |
|---|---|---|
| graphite / 1.cone | 1.3 / 1.2 | 60 |
| graphite / 2.cone | 1.2 / 1.1 | 60 |
| graphite / 3.cone | 1.1 / 1.0 | 80 |
| graphite / cyl. | 1.0 | 90 |
| titanium / cone | 1.1 / 1.0 | 5 |
| titanium / cyl. | 1.0 | 5 |
| iron / cyl. | 1.0 | 10 |
| tungsten / cyl. | 1.0 | 10 |

## 4 Radial heat dissipation for cylindrical collimators

In the equilibrium, the radial heat dissipation through a hollow cylinder with central heating is given by [15]

$$\stackrel{cyl.}{\Rightarrow} \quad \dot{Q} = \frac{\lambda \cdot 2\pi \cdot z \cdot \Delta T}{\ln(r/r_0)} \qquad (2)$$

where $r_0$ is the inner radius of the cylinder, z is the length of the cylinder, and $\lambda$ the heat conductivity. Equation (2) is used to adjust the outer radius of the collimator and the cooling power required to achieve the temperature difference $\Delta T$ between inner and outer surface of the cylinder. The instantaneous heating of the inner part of the cylinder by one bunch train yields maximum temperature rise as shown in Table 2. Between the bunch trains the heat dissipates into the bulk and the temperature at the inner surface of the cylinder decreases. Assuming a homogeneous, radial directed thermal dissipation from the inner hot to the outer cooled surface, the required average cooling power corresponds to $\dot{Q}$. With heat transition coefficients of about 1kW/(m$^2$ K) as typically used for technical cooling solutions, Table 6 presents the power to be discharged per meter of collimator length through the outer surface for the radius r=16cm. The numbers in the table are related to a positron beam polarization of 60%, a bunch train length of about 1ms and a bunch train repetition rate of 5Hz.

The numbers in Table 6 should not hide the fact that the instantaneous power deposition in the photon collimator is huge; it reaches 70kW - 90kW for the parameters considered to achieve 35% - 60% positron polarization.

Tab.6. Average maximal heat flow through a normalized area of about $1m^2$ and $z=1m$ collimator length.

| parts of the 1.collimator $r_0=2.0$mm | heat power through $1m^2$ girthed area [kW/$m^2$] |
| --- | --- |
| py. graphite | 0.30 |
| titanium | 0.01 |
| iron | 0.02 |
| tungsten | 0.03 |
| parts of the 2.collimator $r_0=1.4$mm | heat power through $1m^2$ girthed area [kW/$m^2$] |
| py. graphite | 0.43 |
| titanium | 0.01 |
| iron | 0.02 |
| tungsten | 0.02 |
| parts of the 3.collimator $r_0=1.0$mm | heat power through $1m^2$ girthed area [kW/$m^2$] |
| py. graphite | 0.47 |
| titanium | 0.01 |
| iron | 0.02 |
| tungsten | 0.03 |

**Conclusion**

This multistage collimator system is a proper solution for the photon beam collimation at the ILC positron source. Due to the close correlation between the photon beam intensity, collimation and the degree of polarization the collimator system has to withstand huge heat loads without breakdown during a long operation time. The presented collimator design keeps the material loads in a comfortable regime with an additionally safety margin against failure due to fatique stress. Furthermore, the presented solution can be easily adopted to drive beam energies up to 1TeV.

**Acknowledgments**

We would like to thank the organizers and the host of POSIPOL 2011 for this fruitful and encouraging workshop and for hospitality.
Work supported by the German Federal Ministry of Education and research, Joint Project R&D Accelerator Spin Management, contract No. 05H10GUE.